\documentclass[a4paper, amsfonts, amssymb, amsmath, reprint, showkeys, nofootinbib, twoside,superscriptaddress,citeautoscript,citeautoscript, aps, prl, floatfix]{revtex4-2}

\usepackage{graphicx}
\usepackage{dcolumn}
\usepackage{bm}
\usepackage{amssymb,color}

\begin{document}

\preprint{APS/123-QED}

\title{ \textbf{Mechanism of Shape Symmetry Breaking in Surfactant Mediated Crystal Growth} }

\author{Sam Oaks-Leaf}
\affiliation{\mbox{Department of Chemistry, University of California, Berkeley, CA 94720, USA}}
\author{David T. Limmer}
 \email{Contact author: dlimmer@berkeley.edu}
 \affiliation{\mbox{Department of Chemistry, University of California, Berkeley, CA 94720, USA}}
\affiliation{\mbox{Materials Sciences Division, Lawrence Berkeley National Laboratory, Berkeley, CA 94720, USA}}
\affiliation{\mbox{Chemical Sciences Division, Lawrence Berkeley National Laboratory, Berkeley, CA 94720, USA}}
\affiliation{\mbox{Kavli Energy Nanoscience Institute, Berkeley, CA 94720, USA}}

\date{\today}

\begin{abstract}
 We present a dynamical model of crystal growth, in which it is possible to reliably achieve asymmetric products, beginning from symmetric initial conditions and growing within an isotropic environment. The asymmetric growth is the result of a positive feedback mechanism that amplifies the effect of thermal fluctuations in the coverage of surfactants on the growing crystalline facets. Within our simple model, we are able to understand the kinetic and thermodynamic factors involved in both the onset of symmetry breaking and the persistence of anisotropic growth. 
 We demonstrate that the mechanism is general by studying models with increasing complexity. We argue that this mechanism of symmetry breaking underpins observations of colloidal, seed-mediated syntheses of single crystalline metal nanorods capped with strongly interacting surfactants. The parameters within our model are related to experimental observables such as the concentration, hydrophobicity, and binding strength of the surfactants, which suggests a potential route to optimize the yield of asymmetric products in colloidal nanoparticle syntheses.

\end{abstract}

\maketitle

In crystal growth processes at the nanoscale, it has long been known that surfactants play a key role in determining the shape of the products by altering the growth rates and stability of crystalline facets ~\cite{bakshi_how_2016,song_review_2021}. However, it is only heuristically understood how surfactants can break the symmetry of an initial seed during a crystal growth process~~\cite{liz-marzan_growing_2017, walker_extracting_2023}. A prototypical example of a surfactant-mediated growth process that spontaneously breaks symmetry is the seed-based synthesis of colloidal metal nanorods in solution. Although such syntheses are well-studied experimentally~\cite{jana_wet_2001, wei_seed-mediated_2021,park_growth_2013,tong_control_2017,takenaka_effects_2013,walsh_mechanism_2017, xiao_surfactant_2011, gao_dependence_2003}, there currently does not exist a general theory of crystal growth that can explain the onset of persistent anisotropic growth without asymmetry in the initial conditions or in the environment. This letter presents a simple and general positive feedback mechanism that amplifies thermal fluctuations in the coverage of surfactants and leads to persistent anisotropic growth from symmetric initial conditions.  We show first that this mechanism can be understood within a deterministic model of crystal growth as a dynamical instability~\cite{sekerka_role_1993} in the equations of motion for the average positions of crystalline facets. Then, we demonstrate the necessary conditions on the kinetics of the system to guarantee persistent anisotropic growth in a stochastic system. Finally, within a detailed model, we connect the parameters of the model to experimental observables such as the concentration, hydrophobicity, and binding strength of the surfactants used in the synthetic protocol, opening the door for leveraging insights from the model to optimize the yield of asymmetric products in colloidal nanoparticle syntheses. 

The thermodynamically preferred shape of a crystal at any size is determined by the Wulff construction~\cite{marks_nanoparticle_2016} and contains the same symmetry as the underlying lattice. Molecular simulations have demonstrated the stability of surfactant micelles on different crystalline facets~\cite{da_silva_applying_2025}, and while this can alter the relative stabilities of facets it does not explain asymmetric growth.
Kinetic generalizations of a Wulff construction that express the shape as ratios of crystal growth rates also predict isotropic crystals in isotropic environments provided symmetric seeds. If surfactants modify the growth rates of all equivalent facets in the same way, the symmetry of the initial seed will be preserved~\cite{bassani_kinetically_2025, balankura_predicting_2016, seyed-razavi_surface_2011, xia_shape-controlled_2015}. Therefore, a theory that can explain spontaneous symmetry breaking in a crystal growth process must include thermal fluctuations as well as a mechanism for these fluctuations to persist. 

The simplest model capable of exhibiting symmetry breaking during growth is one in which two equivalent facets grow in perpendicular directions. We initially consider growth in two dimensions on a square lattice, with the average position of each facet, $h_x$ and $h_y$ (\textbf{Fig. 1}). We suppose growth is driven by a constant chemical potential difference, $\Delta\mu_g$, between monomers in a fluid and the crystal, and that the temperature is much lower than the binding strength between monomers such that growth on a given crystalline facet proceeds only via the nucleation of complete layers. The simplest possible equations of motion that are thermodynamically consistent with such a process are
\begin{equation}
    \frac{\mathrm{d}h_i}{\mathrm{d}t} = K_g^{+}(1-e^{-\beta(\Delta\mu_g h_j-\epsilon_g)})
    \label{symDet}
\end{equation}
where $i$ and $j$ represent $x$ and $y$ symmetrically, $\beta$ is the inverse temperature times Boltzmann's constant, $\epsilon_g$ is the energetic cost of breaking one bond between a pair of metal atoms, and  $K_g^{+}$ is the average growth rate of a new facet. We note that in principle $K_g^{+}$ could 
depend on $h_j$, but it is not necessary to consider this additional detail at present. Importantly, just from thermodynamic considerations, the evolution of $h_x$ and $h_y$ are coupled and in such a way that could never lead to asymmetric growth. If $h_x > h_y$, then clearly $\mathrm{d}h_x/\mathrm{d}t \leq \mathrm{d}h_y/\mathrm{d}t$. 

\begin{figure}[t]
\includegraphics[width=0.45\textwidth]{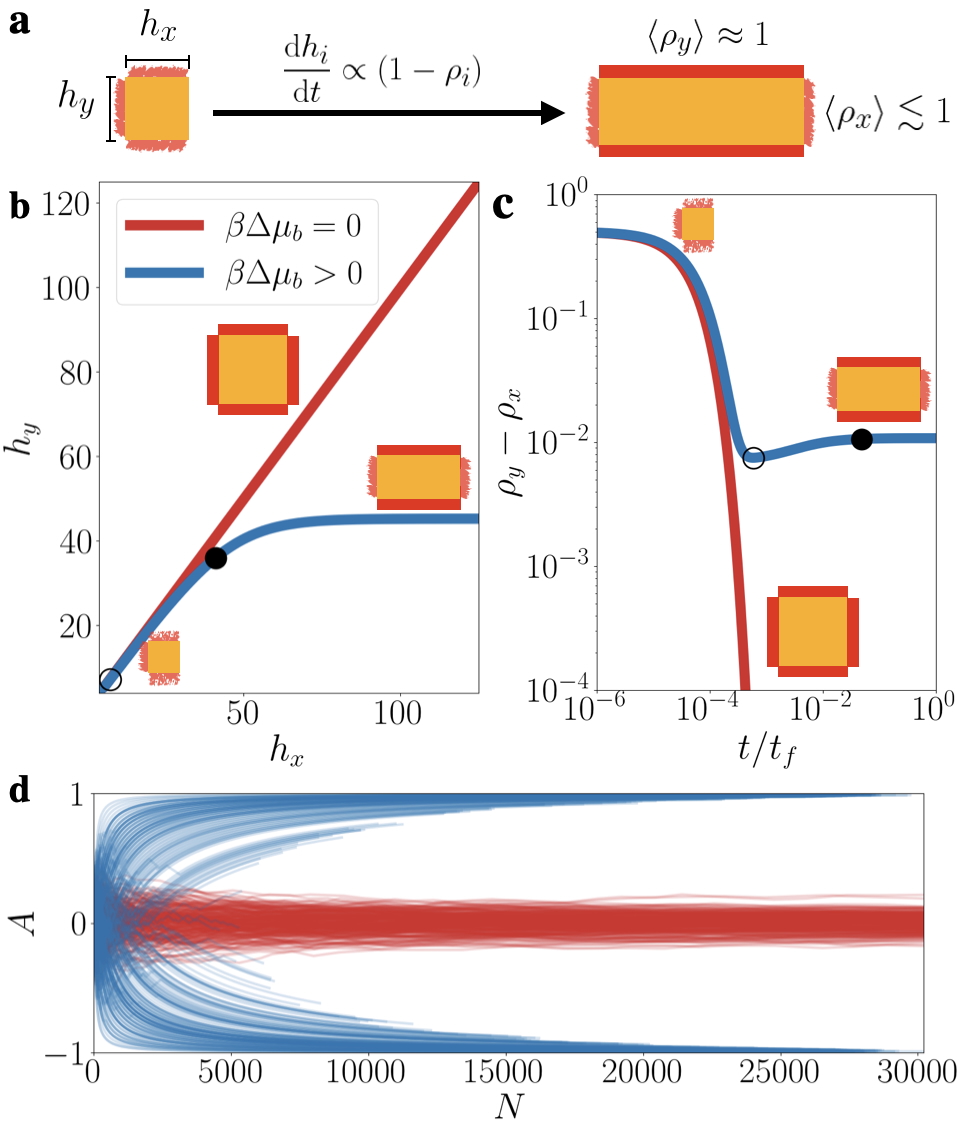}
\caption{\label{fig:fig1} 
Surfactants dynamically stabilize anisotropic growth. \textbf{a}, Schematic of the mechanism of symmetry breaking in a square lattice with red being the surfactant and yellow the metal nanocrystal. \textbf{b}, Parametric plot of average facet positions, $h_x$ and $h_y$, evolving under Eq.~\ref{blockDet}, with $\epsilon_g=\epsilon_b=0$ and $\beta\Delta\mu_g = 1.0$ The label, $\beta\Delta\mu_b>0.0$, corresponds to $\beta\Delta\mu_b=0.1$. Trajectories begin from initial condition $h_x = h_y = 5.0$, $\rho_x = 0.0$, and $\rho_y = 0.5$. $K_b^{+}$ is equal to $K_g^{+}$ and $t_fK_g^{+}= 10^{4}$. \textbf{c}, The same trajectories as in (\textbf{b}) but showing the average difference in surfactant densities between the $y$ and $x$ facets as a function of time. Circular markers correspond to the same points in time as the corresponding markers in (\textbf{b}). \textbf{d}, Trajectories sampled from stochastic coarse-grained model with KMC. The initial condition is totally symmetric, with $h_x = h_y = 5$ and $\rho_x = \rho_y = 1$. All other parameters are the same as in (\textbf{b}) and (\textbf{c}).}
\end{figure}

In practice, the growth of nanocrystals from solution occurs in the presence of passivating, surface-active ligands. To include such an effect, we introduce $\rho_i$ which varies between 0 and 1, as the average surfactant density on the $i$th facet. We ignore the spatial dependence of the surfactant density, assuming that the surfactant molecules interact strongly enough that the only thermodynamically favorable states are a fully covered or an empty facet. We first assume that a covered facet cannot grow, so that the evolution equations become
\begin{eqnarray}
    \frac{\mathrm{d}h_i}{\mathrm{d}t} &= (1 - \rho_i)K_g^{+}(1-e^{-\beta(\Delta\mu_g h_j-\epsilon_g)}) \nonumber \\
    \frac{\mathrm{d}\rho_i}{\mathrm{d}t} &= K_b^{+} \big(1 - \rho_i(1 +e^{-\beta(\Delta\mu_b h_j-\epsilon_b)})\big) 
    \label{blockDet}
\end{eqnarray}
where we have introduced the subscript, $b$, for each parameter associated with the ligand species as they ``block" crystal growth. If growth is thermodynamically preferred for facet $i$, $\Delta\mu_g h_j>\epsilon$, then its growth rate will be greater than or equal to zero. A rate of zero growth is only achieved if $\rho_i = 1$. 

From Eq.~\ref{blockDet}, provided some initial infinitesimal asymmetry and $\beta\Delta\mu_b > 0$, this model predicts persistent asymmetric growth via a positive reinforcement mechanism. The mechanism is illustrated schematically in \textbf{Fig. 1a} and an example of asymmetric growth is shown in \textbf{Fig. 1b and c}. A system initialized as a perfect square with symmetric surfactant coverage will grow as a square indefinitely. However, if the surfactant coverage is perturbed away from this symmetry point, such that for instance, $\rho_y > \rho_x$, then a positive chemical driving force for surfactant binding will lead to a quasi-steady state for the surfactant densities in which $\rho_y$ remains greater than $\rho_x$. This occurs because at the moment of the perturbation, $\mathrm{d}h_x/\mathrm{d}t$ becomes greater than $\mathrm{d}h_y/\mathrm{d}t$. Therefore, $h_x$ becomes slightly larger than $h_y$, and in turn $\mathrm{d}\rho_y/\mathrm{d}t$ becomes larger than $\mathrm{d}\rho_x/\mathrm{d}t$, generating a positive feedback process. As shown in \textbf{Fig. 1 b and c}, if $\beta\Delta\mu_b=0$, this feedback does not exist and the perturbation will be quickly corrected so that in the long time limit $\rho_x = \rho_y$ and $\mathrm{d}h_x/\mathrm{d}t = \mathrm{d}h_y/\mathrm{d}t$.

This feedback mechanism can be understood by invoking a quasi-steady-state assumption for the surfactants, $\mathrm{d} \rho_i/\mathrm{d}t=0$. Transforming coordinates to $N=h_xh_y$, a measure of the size of the system, and $A=(h_x - h_y)/(h_x + h_y)$, a measure of the asymmetry, the corresponding equations of motion are
\begin{eqnarray}
    \frac{\mathrm{d}A}{\mathrm{d}t} &=& K_g^{+}\frac{1-A^2}{\sqrt{N}}\bigg[Q\bigg(\frac{1-A}{1+A}\bigg) - Q\bigg(\frac{1+A}{1-A}\bigg)\bigg]
\label{AandN}
\end{eqnarray}
with 
\begin{equation}
Q(x) = \frac{1-\exp(-\Delta\mu_g\sqrt{Nx}+\epsilon_g)}{1+\exp(\Delta\mu_b\sqrt{Nx}-\epsilon_b)}x
\end{equation}
If $\Delta\mu_b>0$, then in the large $N$ limit, $Q$ goes to zero for any $x$ and thus so does $\mathrm{d}A/\mathrm{d}t$, as would be expected due to the blocking action of the surfactants. However, if $A=\pm \delta$, with $\delta$ any finite value, then in the large $N$ limit $A$ and $\mathrm{d}A/\mathrm{d}t$ will have the same sign. This can be seen by considering that if $0<A<1$ then $(1-A)/(1+A)<(1+A)/(1-A)$. In the large $N$ limit, $Q$ is a decreasing function and therefore $\mathrm{d}A/\mathrm{d}t$ is positive. This implies that if $A$ is at any time perturbed above zero, the asymptotic value of $A$ will be $1$. Similarly, if $A$ is perturbed below zero, its asymptotic value will be $-1$. In this most simple system of surfactant-mediated growth, the asymptotic product will always be maximally asymmetric, provided that the system continues to grow.

The deterministic model amplifies any asymmetric perturbation, but it cannot describe spontaneous symmetry breaking because it does not contain thermal fluctuations. To see true symmetry breaking we construct a discrete Markov model with rates set such that they are described by Eq.~\eqref{blockDet} in a mean-field limit. Our state space is described by vectors $|h_x, h_y, \rho_x, \rho_y\rangle$ where  $h_x,h_y \in \mathbb{N}$ and $\rho_x, \rho_y \in \{0, 1\}$. We see in \textbf{Fig. 1d} that in this model, simulated with kinetic Monte Carlo (KMC)\cite{gillespie_stochastic_2007}, if $\Delta\mu_b > 0$, we can begin with a square seed with symmetric surfactant coverage and reliably tend toward a totally asymmetric product. Thermal fluctuations are equally likely to increase or decrease $A$ so that the system is equally likely to relax toward $A =\pm 1$, but all trajectories become maximally asymmetric in the limit that time goes to infinity. 

\begin{figure}[t]
\includegraphics[width=0.45\textwidth]{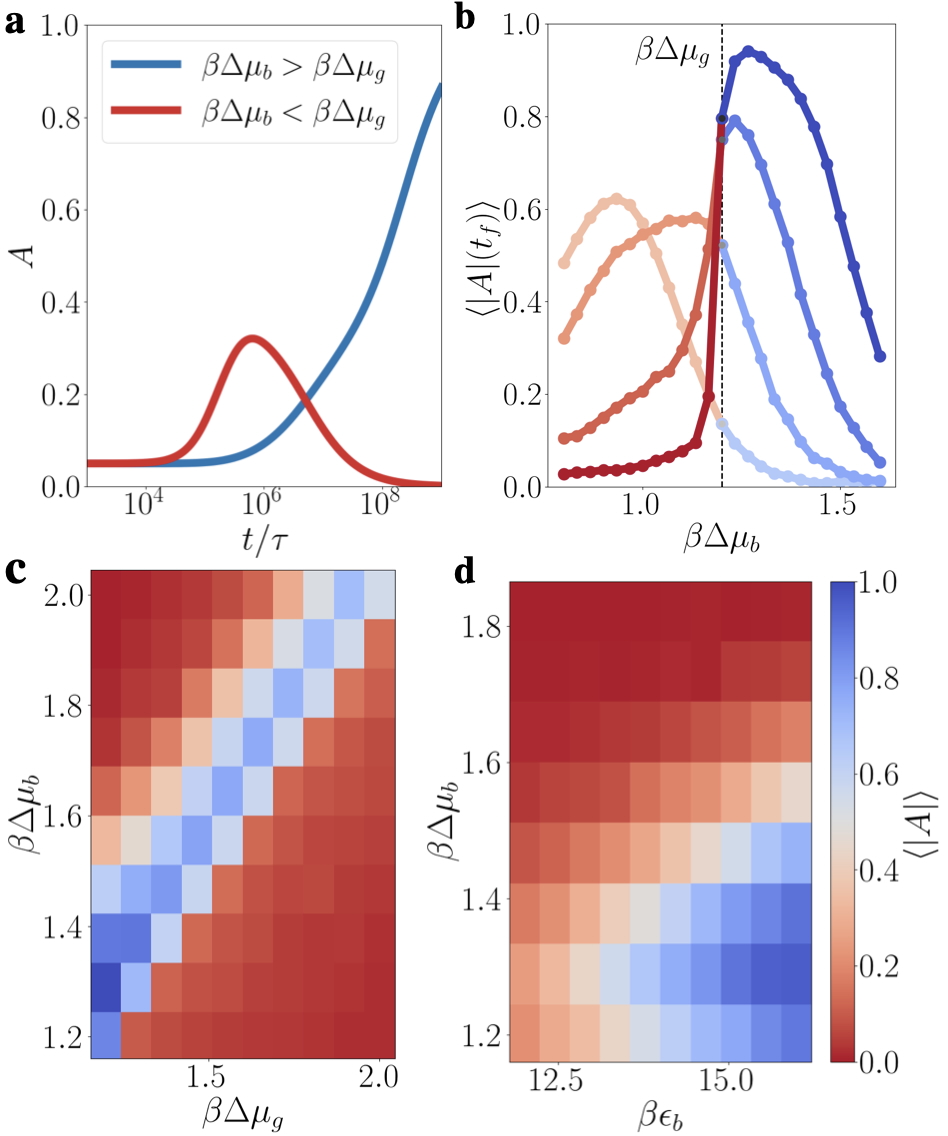}
\caption{Competitive growth processes leads to a regime of transient symmetry breaking and a regime of persistent anisotropic growth. \textbf{a}, Mean-field trajectories of the asymmetry and total size under conditions where $\beta\Delta\mu_b = 1.0$ and $\beta\Delta\mu_b = 1.3$. $\tau$ is the inverse of the typical attachment frequency for both species (\textbf{Supplementary} \textbf{Material}).  $\beta\epsilon_g = 12.5$, $\beta\epsilon_b = 15.0$, and $\beta\Delta\mu_g=1.2$ in both (\textbf{a}) and (\textbf{b}). For the trajectories of (\textbf{a}), the initial condition is $A=0.05$, $N=256$, and total surfactant coverage. \textbf{b}, Average of the absolute value of the asymmetry as a function of $\Delta\mu_b$, taken from KMC trajectories at final times $t_f/\tau = 10^6,\hspace{0.5mm}10^7,\hspace{0.5mm}10^8,$ and $10^9$. The initial condition is totally symmetric with $h_i=16$ and $\rho_i^{a}=1$. \textbf{c} and \textbf{d}, two dimensional phase diagrams of the average of the absolute value of kMC trajectories taken at a final time of $t_f/\tau=5\times 10^8$. Initial conditions are the same as in (\textbf{b}).}
  \label{fig:Fig2}
\end{figure}

The model we have presented so far identifies a mechanism for persistent asymmetric growth but is idealized. One major assumption we have made is that the only way to grow a new crystal layer on a facet covered by surfactant is to remove the surfactant and then nucleate a new layer. However, in reality there could be other growth mechanisms. Particularly, one can imagine a process of competitive binding in which surfactant molecules are removed and immediately replaced by metal adatoms.  We can add such a competitive growth process to our model and test the robustness of our initial observations to it.

The competitive process that grows the nanocrystal in the presence of surfactants will have a thermodynamic driving dependent on the difference between the two chemical potentials for the two species. Within the deterministic framework we can add it to our equations of motion as,
\begin{eqnarray}
    \frac{\mathrm{d}h_i}{\mathrm{d}t} &=& (1 - \rho_i)\left [K_g^{+}(1-f_{g}^j) - K_{bg}\frac{f_{g}^j}{f_{b}^j}\right ] + \rho_i K_{bg} \nonumber \\  
    \frac{\mathrm{d}\rho_i}{\mathrm{d}t} &=& (1-\rho_i)\left [ K_b^{+} +K_{bg}\frac{f_{g}^j}{f_{b}^j}\right ] -\rho_i\big(K_b^{+}f_{b}^j + K_{bg}\big)
\label{kTilde}
\end{eqnarray}
where $K_{bg}$ is an effective rate for the competitive growth process, and $f_{a}^l=\exp(-\beta\Delta\mu_a h_l + \beta\epsilon_a)$. 

As we add the competitive growth process, we also relax several assumptions within our initial model. We allow the values of $K_g^{+}$, $K_b^{+}$ to depend on the length of the facet as well as the thermodynamic parameters and choose the forms of $K_{b}^{+}$ and $K_{g}^{+}$ such that they agree with an analytical form for a nucleation process of each species in one dimension~\cite{joswiak_critical_2016} (see \textbf{Supplementary Material}). We then evaluate the effective rate $K_{bg}$ as a function of $K_g^{+}$ and $K_b^{+}$ such that it describes the competitive process in which surfactant molecules are removed from the edge of a micelle and immediately replaced with metal adatoms (\textbf{Supplementary Material}). Finally, we also choose to represent all four facets of the square lattice. Thus, our state space is expanded to vectors of the form $|h_x, h_y, \rho^{u}_x, \rho^{l}_x, \rho_y^{u}, \rho_y^{l}\rangle$. The four surfactant coverages obey the same dynamics as the original two, and $h_i$ can now decrease and increase in two directions depending on the values of $\rho_i^{u}$ and $\rho_i^{l}$. 

Upon adding these complications to the model, the mechanism of symmetry breaking remains. However, the addition of the competitive growth process shifts the dynamical phase diagram for the asymmetry. In both the mean field limit and the stochastic simulation, the competitive process creates a regime of transient asymmetric growth when $0 < \beta\Delta\mu_b < \beta\Delta\mu_g$ and a regime of persistent asymmetric growth when $\beta\Delta\mu_b > \beta\Delta\mu_g$. In the transient regime, an asymmetric perturbation will be initially amplified but decays in the long-time limit (\textbf{Fig. 2a}). The details of this transient behavior depend on all parameters and the specific forms of $K_b^{+}$, $K_g^{+}$, and $K_{bg}$. However, the asymptotic behavior of this model, in the long time or equivalently the large $N$ limit, depends only on whether $\Delta\mu_b$ is greater or less than $\Delta\mu_g$. As long as $K_{bg}$ is finite, the asymmetry in the transient regime will eventually decay to zero as the asymmetry in the persistent regime tends to its maximal value. For a finite growth time, the value of the average asymmetry at the final time will depend strongly on the value of $\Delta\mu_b$ relative to $\Delta\mu_g$. For short times, maximal asymmetry will be attained within the transient regime where $\Delta\mu_b < \Delta\mu_g$, as larger values of $\Delta\mu_b$ will prevent the system from growing quickly enough to generate significant asymmetry (\textbf{Fig. 2b}).

In \textbf{Figures 2c and d} we fix a final time at which much of the transient asymmetry has decayed and study the dependence of the asymmetry on the chemical driving forces for each species and the surface tension term of the surfactants, $\epsilon_b$. We plot the average absolute value of asymmetry at the final time as a function of $\beta\Delta\mu_g$ and $\beta\Delta\mu_b$ and see there exists an approximately linear band along the line where $\beta\Delta\mu_b$ is just greater than $\beta\Delta\mu_g$ at which maximal asymmetry can be achieved (\textbf{Fig. 2c}). We also observe that the final asymmetry is an increasing function of $\epsilon_b$ regardless of the chemical driving force. This is expected if one considers the expression for the steady-state, average surfactant density, $\langle \rho_x \rangle = \exp(\beta\Delta\mu_b h_y)/\big(\exp(\beta\Delta\epsilon_b) + \exp(\beta\Delta\mu_b h_y)\big)$. If $h_y > h_x$, $\langle \rho_x \rangle / \langle\rho_y\rangle$ is an increasing function of $\epsilon_b$ and its magnitude directly determines the degree of asymmetry.

\begin{figure}[t]
\includegraphics[width=0.45\textwidth]{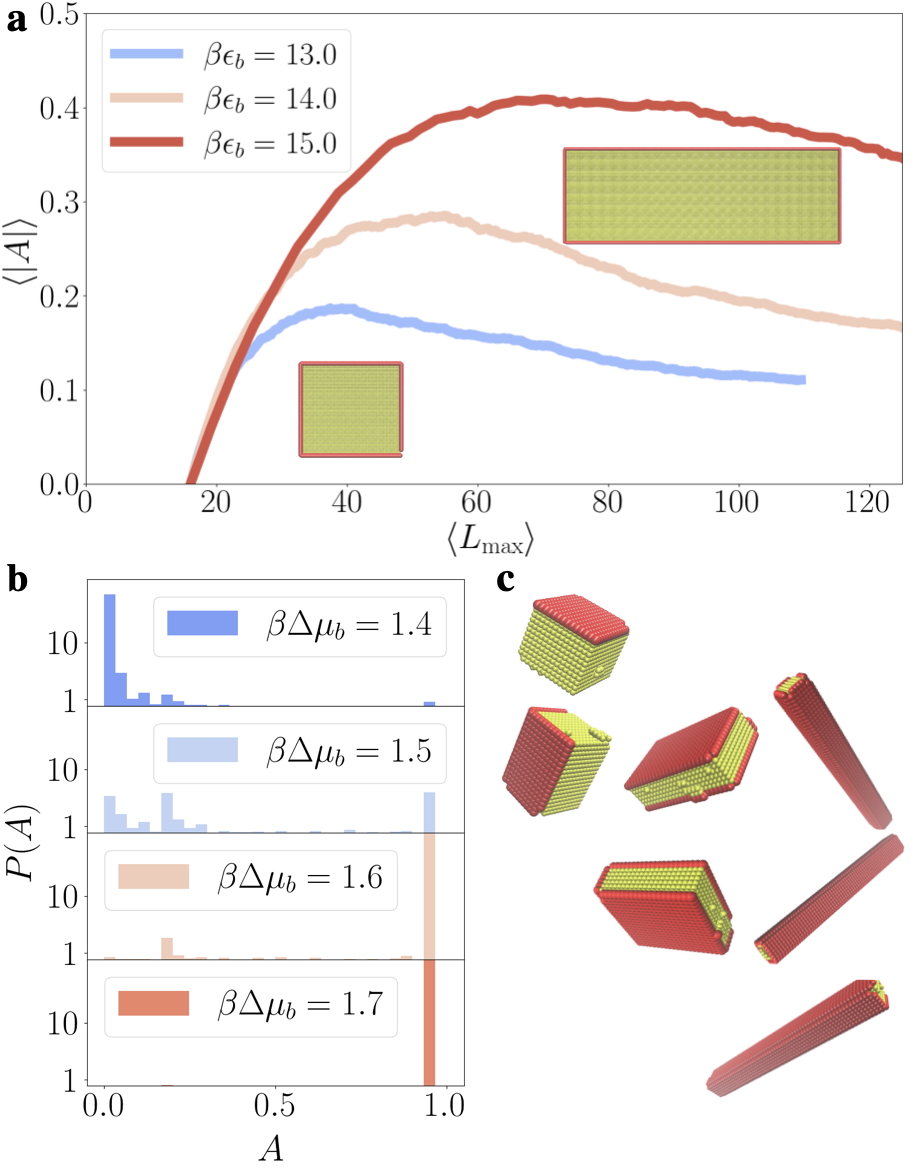}
\caption{Symmetry breaking growth in a two component lattice gas. \textbf{a}, Parametric plots of $\langle |A|\rangle$ versus the average of the length of the longest facet of the growing crystal, $\langle L_\mathrm{max}\rangle$. Trajectories sampled from KMC with $\beta\Delta\mu_b=1.7$, $\beta\Delta\mu_g=1.5$ and $\beta\epsilon_g=12.5$. Initial condition is a perfect square of side length $16$ atoms and complete surfactant coverage. \textbf{b}, Probability, $P(A)$, of a final asymmetry after growing in the three dimensional model from a 216 atom cube with complete surfactant coverage to a product with 2000 atoms. $\beta\epsilon_b=\beta\epsilon_g=5.0$, with $\beta\Delta\mu_g=2.0$ All ensembles are for 1024 independent trajectories. \textbf{c}, Representative configurations from each the histograms of (\textbf{b}).
  \label{fig:Fig3}}
\end{figure}

To demonstrate the generality of the positive feedback mechanism, we consider a two component lattice gas. Such a model relaxes the assumptions of uniform surfactant coverage and perfectly faceted surfaces. One species represents the nanoparticle species, $g$, and a surfactant species, $b$, just as in the previous models. While still a simplified description of atomistic crystal growth, lattice gases are known to give qualitatively correct predictions particularly in thermodynamically governed transformations of metal nanocrystals~\cite{ye_single-particle_2016, lai_nucleation-mediated_2023, lai_reshaping_2019, han_real-time_2014, turner_modeling_2016}. Each species has a binding strength for only its own type, $\epsilon_g$ and $\epsilon_b$ respectively, as well as a chemical potential, $\mu_g$ and $\mu_b$, all of which enter into the lattice gas Hamiltonian\cite{chandler_introduction_1988}. We define chemical potential differences for each species such that $\Delta\mu_g = \mu_g + \epsilon_g z/2$ and $\Delta\mu_b = \mu_b +\epsilon_b z/4$, where $z$ is the coordination number of the lattice. The growth of a single crystal is enforced by restricting the rates such that attachment of either species is allowed only at unfilled sites that neighbor a site filled by species $g$. Detachment of either species is also disallowed if it would create an isolated filled site. To capture the blocking action of the surfactants, we enforce that a species of type $b$ may only have up to $z-1$ total neighbors and only $z/2$ neighbors of type $b$. These constraints ensure blocking species remain on the surface of the nanocrystal throughout growth.

We begin a growth process with a symmetric seed composed of the nanoparticle species and with surfaces covered in the blocking species. We propagate the system forward in time using KMC\cite{gillespie_stochastic_2007}, with rates of attachment and detachment for each species that obey local detailed balance with respect to the lattice gas Hamiltonian. We observe that the same positive feedback mechanism occurs in the lattice gas and that the degree of asymmetry monotonically increases with $\epsilon_b$ (\textbf{Fig. 3a}). Interestingly, even when $\Delta\mu_b >\Delta\mu_g$, the asymmetry in a two-dimensional lattice gas model is transient. This is because the facets of the two-dimensional crystal are themselves one-dimensional. An assumption of our previous analysis was that the attraction between bound surfactants is strong enough that they are below a critical condensation temperature and thus the facet exists predominately in either a covered or uncovered state. In one dimension there is no such finite critical temperature, and thus for any value of $\beta\epsilon_b$ there will be a length of the crystal facet where a supercritical state becomes entropically favorable. At this point, the positive feedback mechanism no longer exists and the asymmetry will decay on average (\textbf{Fig. 3a}). 

In the growth of a three-dimensional simple cubic crystal, we can set $\beta\epsilon_b$ such that the surfactants are below the critical temperature of a two-dimensional square lattice gas. Doing so, we see that the positive feedback mechanism reliably occurs and we do not observe a decay of the asymmetry, at least within the timescales we are able to simulate. Additionally, we observe that in three-dimensional growth, one can find values of $\Delta\mu_b$ in which it is likely to preferentially grow in two of the three equivalent directions rather than just one, forming platelets rather than rods. At sufficiently small values of $\Delta\mu_b$ relative to $\Delta\mu_g$, the surfactants are only weakly bound and the final products are likely to be nearly perfectly symmetric. Increasing $\Delta\mu_b$ one observes a regime in which there are significant populations of platelets within the final products. These are crystals in which only one of the three equivalent dimensions remains near its initial size throughout the growth process. Platelets form as a result of a fluctuation in the initial stages of growth, in which two facets are uncovered at once. At large enough $\Delta\mu_b$, this simultaneous fluctuation becomes very unlikely, and one finds only a single population of rod-like products. This trend is quantified in \textbf{Fig. 3b} with the definition of asymmetry in three dimensions taken as the relative shape anisotropy, $A=\frac{1}{2}(3\frac{h_x^4 + h_y^4 + h_z^4}{(h_x^2 +h_y^2 +h_z^2)^2}-1)$\cite{theodorou_shape_1985}.

We have identified a general mechanism of spontaneous shape symmetry breaking in crystal growth processes mediated by surfactants. We have demonstrated the robustness of this mechanism in a series of models of crystal growth of increasing complexity. The mechanism relies on two basic thermodynamic requirements. The surfactants must have an effective attraction strong enough that they are below a critical condensation temperature, and there must be a finite chemical driving force for the surfactants to bind to the crystal such that the rate to remove a surfactant layer from a facet decreases as the size of the facet increases. 

In a colloidal nanocrystal synthesis, the effective attraction between surfactants could come from a hydrophobic interaction between their aliphatic tails. The chemical driving force is the reversible work to move a surfactant molecule from the nanocrystal surface to the solution.
Exact values for these energies would be difficult to compute in a realistic system, but we should expect that the hydrophobic attraction would increase with the length of the surfactant tail~\cite{chandler_interfaces_2005} and that the reversible work would increase with the concentration of surfactants in solution. In the colloidal synthesis of gold nanorods it has been demonstrated experimentally that increasing the length of the surfactant tail and increasing the surfactant concentration both lead to higher final aspect ratios on average~\cite{takenaka_effects_2013, gao_dependence_2003}, consistent with our mechanism. It is straightforward to generalize the models we have presented to more relevant crystal structures and to include more complicated interactions between both pairs of surfactant molecules and between a surfactant molecule and the nanocrystal surface. With these modifications, we may be able to use the model as a predictive tool to optimize synthetic protocols of anisotropic metal nanocrystals.

The code and data that support these findings are openly available at \cite{oaksleaf_data}

\emph{Awknowledgements} This work was supported by the U.S. Department of
Energy, Office of Science, Office of Basic Energy Sciences, Materials Sciences and Engineering Division under Contract No. DE-AC02-05-CH11231 within
the in-situ TEM program (KC22ZH). Work at the Molecular Foundry of Lawrence Berkeley National Laboratory was supported by the U.S. Department of Energy under Contract No. DE-AC02-05CH11231. 

\bibliography{main}

\begin{thebibliography}{30}%
\makeatletter
\providecommand \@ifxundefined [1]{%
 \@ifx{#1\undefined}
}%
\providecommand \@ifnum [1]{%
 \ifnum #1\expandafter \@firstoftwo
 \else \expandafter \@secondoftwo
 \fi
}%
\providecommand \@ifx [1]{%
 \ifx #1\expandafter \@firstoftwo
 \else \expandafter \@secondoftwo
 \fi
}%
\providecommand \natexlab [1]{#1}%
\providecommand \enquote  [1]{``#1''}%
\providecommand \bibnamefont  [1]{#1}%
\providecommand \bibfnamefont [1]{#1}%
\providecommand \citenamefont [1]{#1}%
\providecommand \href@noop [0]{\@secondoftwo}%
\providecommand \href [0]{\begingroup \@sanitize@url \@href}%
\providecommand \@href[1]{\@@startlink{#1}\@@href}%
\providecommand \@@href[1]{\endgroup#1\@@endlink}%
\providecommand \@sanitize@url [0]{\catcode `\\12\catcode `\$12\catcode `\&12\catcode `\#12\catcode `\^12\catcode `\_12\catcode `\%12\relax}%
\providecommand \@@startlink[1]{}%
\providecommand \@@endlink[0]{}%
\providecommand \url  [0]{\begingroup\@sanitize@url \@url }%
\providecommand \@url [1]{\endgroup\@href {#1}{\urlprefix }}%
\providecommand \urlprefix  [0]{URL }%
\providecommand \Eprint [0]{\href }%
\providecommand \doibase [0]{https://doi.org/}%
\providecommand \selectlanguage [0]{\@gobble}%
\providecommand \bibinfo  [0]{\@secondoftwo}%
\providecommand \bibfield  [0]{\@secondoftwo}%
\providecommand \translation [1]{[#1]}%
\providecommand \BibitemOpen [0]{}%
\providecommand \bibitemStop [0]{}%
\providecommand \bibitemNoStop [0]{.\EOS\space}%
\providecommand \EOS [0]{\spacefactor3000\relax}%
\providecommand \BibitemShut  [1]{\csname bibitem#1\endcsname}%
\let\auto@bib@innerbib\@empty
\bibitem [{\citenamefont {Bakshi}(2016)}]{bakshi_how_2016}%
  \BibitemOpen
  \bibfield  {author} {\bibinfo {author} {\bibfnamefont {M.~S.}\ \bibnamefont {Bakshi}},\ }\bibfield  {title} {\bibinfo {title} {How {Surfactants} {Control} {Crystal} {Growth} of {Nanomaterials}},\ }\href {https://doi.org/10.1021/acs.cgd.5b01465} {\bibfield  {journal} {\bibinfo  {journal} {Crystal Growth \& Design}\ }\textbf {\bibinfo {volume} {16}},\ \bibinfo {pages} {1104} (\bibinfo {year} {2016})},\ \bibinfo {note} {publisher: American Chemical Society}\BibitemShut {NoStop}%
\bibitem [{\citenamefont {Song}\ \emph {et~al.}(2021)\citenamefont {Song}, \citenamefont {Gao}, \citenamefont {Guo}, \citenamefont {Zhang}, \citenamefont {Li}, \citenamefont {You},\ and\ \citenamefont {Du}}]{song_review_2021}%
  \BibitemOpen
  \bibfield  {author} {\bibinfo {author} {\bibfnamefont {T.}~\bibnamefont {Song}}, \bibinfo {author} {\bibfnamefont {F.}~\bibnamefont {Gao}}, \bibinfo {author} {\bibfnamefont {S.}~\bibnamefont {Guo}}, \bibinfo {author} {\bibfnamefont {Y.}~\bibnamefont {Zhang}}, \bibinfo {author} {\bibfnamefont {S.}~\bibnamefont {Li}}, \bibinfo {author} {\bibfnamefont {H.}~\bibnamefont {You}},\ and\ \bibinfo {author} {\bibfnamefont {Y.}~\bibnamefont {Du}},\ }\bibfield  {title} {\bibinfo {title} {A review of the role and mechanism of surfactants in the morphology control of metal nanoparticles},\ }\href {https://doi.org/10.1039/D0NR07339C} {\bibfield  {journal} {\bibinfo  {journal} {Nanoscale}\ }\textbf {\bibinfo {volume} {13}},\ \bibinfo {pages} {3895} (\bibinfo {year} {2021})},\ \bibinfo {note} {publisher: Royal Society of Chemistry}\BibitemShut {NoStop}%
\bibitem [{\citenamefont {Liz-Marzán}\ and\ \citenamefont {Grzelczak}(2017)}]{liz-marzan_growing_2017}%
  \BibitemOpen
  \bibfield  {author} {\bibinfo {author} {\bibfnamefont {L.~M.}\ \bibnamefont {Liz-Marzán}}\ and\ \bibinfo {author} {\bibfnamefont {M.}~\bibnamefont {Grzelczak}},\ }\bibfield  {title} {\bibinfo {title} {Growing anisotropic crystals at the nanoscale},\ }\href {https://doi.org/10.1126/science.aam8774} {\bibfield  {journal} {\bibinfo  {journal} {Science}\ }\textbf {\bibinfo {volume} {356}},\ \bibinfo {pages} {1120} (\bibinfo {year} {2017})},\ \bibinfo {note} {publisher: American Association for the Advancement of Science}\BibitemShut {NoStop}%
\bibitem [{\citenamefont {Walker}\ \emph {et~al.}(2023)\citenamefont {Walker}, \citenamefont {Lee}, \citenamefont {Dagdelen}, \citenamefont {Cruse}, \citenamefont {Gleason}, \citenamefont {Dunn}, \citenamefont {Ceder}, \citenamefont {Paul Alivisatos}, \citenamefont {A. Persson},\ and\ \citenamefont {Jain}}]{walker_extracting_2023}%
  \BibitemOpen
  \bibfield  {author} {\bibinfo {author} {\bibfnamefont {N.}~\bibnamefont {Walker}}, \bibinfo {author} {\bibfnamefont {S.}~\bibnamefont {Lee}}, \bibinfo {author} {\bibfnamefont {J.}~\bibnamefont {Dagdelen}}, \bibinfo {author} {\bibfnamefont {K.}~\bibnamefont {Cruse}}, \bibinfo {author} {\bibfnamefont {S.}~\bibnamefont {Gleason}}, \bibinfo {author} {\bibfnamefont {A.}~\bibnamefont {Dunn}}, \bibinfo {author} {\bibfnamefont {G.}~\bibnamefont {Ceder}}, \bibinfo {author} {\bibfnamefont {A.}~\bibnamefont {Paul Alivisatos}}, \bibinfo {author} {\bibfnamefont {K.}~\bibnamefont {A. Persson}},\ and\ \bibinfo {author} {\bibfnamefont {A.}~\bibnamefont {Jain}},\ }\bibfield  {title} {\bibinfo {title} {Extracting structured seed-mediated gold nanorod growth procedures from scientific text with {LLMs}},\ }\href {https://doi.org/10.1039/D3DD00019B} {\bibfield  {journal} {\bibinfo  {journal} {Digital Discovery}\ }\textbf {\bibinfo {volume} {2}},\ \bibinfo {pages} {1768} (\bibinfo {year} {2023})},\ \bibinfo {note} {publisher:
  Royal Society of Chemistry}\BibitemShut {NoStop}%
\bibitem [{\citenamefont {Jana}\ \emph {et~al.}(2001)\citenamefont {Jana}, \citenamefont {Gearheart},\ and\ \citenamefont {Murphy}}]{jana_wet_2001}%
  \BibitemOpen
  \bibfield  {author} {\bibinfo {author} {\bibfnamefont {N.~R.}\ \bibnamefont {Jana}}, \bibinfo {author} {\bibfnamefont {L.}~\bibnamefont {Gearheart}},\ and\ \bibinfo {author} {\bibfnamefont {C.~J.}\ \bibnamefont {Murphy}},\ }\bibfield  {title} {\bibinfo {title} {Wet {Chemical} {Synthesis} of {High} {Aspect} {Ratio} {Cylindrical} {Gold} {Nanorods}},\ }\href {https://doi.org/10.1021/jp0107964} {\bibfield  {journal} {\bibinfo  {journal} {J. Phys. Chem. B}\ }\textbf {\bibinfo {volume} {105}},\ \bibinfo {pages} {4065} (\bibinfo {year} {2001})},\ \bibinfo {note} {publisher: American Chemical Society}\BibitemShut {NoStop}%
\bibitem [{\citenamefont {Wei}\ \emph {et~al.}(2021)\citenamefont {Wei}, \citenamefont {Deng}, \citenamefont {Zhang}, \citenamefont {Cheng},\ and\ \citenamefont {Li}}]{wei_seed-mediated_2021}%
  \BibitemOpen
  \bibfield  {author} {\bibinfo {author} {\bibfnamefont {M.-Z.}\ \bibnamefont {Wei}}, \bibinfo {author} {\bibfnamefont {T.-S.}\ \bibnamefont {Deng}}, \bibinfo {author} {\bibfnamefont {Q.}~\bibnamefont {Zhang}}, \bibinfo {author} {\bibfnamefont {Z.}~\bibnamefont {Cheng}},\ and\ \bibinfo {author} {\bibfnamefont {S.}~\bibnamefont {Li}},\ }\bibfield  {title} {\bibinfo {title} {Seed-{Mediated} {Synthesis} of {Gold} {Nanorods} at {Low} {Concentrations} of {CTAB}},\ }\href {https://doi.org/10.1021/acsomega.1c00510} {\bibfield  {journal} {\bibinfo  {journal} {ACS Omega}\ }\textbf {\bibinfo {volume} {6}},\ \bibinfo {pages} {9188} (\bibinfo {year} {2021})},\ \bibinfo {note} {publisher: American Chemical Society}\BibitemShut {NoStop}%
\bibitem [{\citenamefont {Park}\ \emph {et~al.}(2013)\citenamefont {Park}, \citenamefont {Drummy}, \citenamefont {Wadams}, \citenamefont {Koerner}, \citenamefont {Nepal}, \citenamefont {Fabris},\ and\ \citenamefont {Vaia}}]{park_growth_2013}%
  \BibitemOpen
  \bibfield  {author} {\bibinfo {author} {\bibfnamefont {K.}~\bibnamefont {Park}}, \bibinfo {author} {\bibfnamefont {L.~F.}\ \bibnamefont {Drummy}}, \bibinfo {author} {\bibfnamefont {R.~C.}\ \bibnamefont {Wadams}}, \bibinfo {author} {\bibfnamefont {H.}~\bibnamefont {Koerner}}, \bibinfo {author} {\bibfnamefont {D.}~\bibnamefont {Nepal}}, \bibinfo {author} {\bibfnamefont {L.}~\bibnamefont {Fabris}},\ and\ \bibinfo {author} {\bibfnamefont {R.~A.}\ \bibnamefont {Vaia}},\ }\bibfield  {title} {\bibinfo {title} {Growth {Mechanism} of {Gold} {Nanorods}},\ }\href {https://doi.org/10.1021/cm303659q} {\bibfield  {journal} {\bibinfo  {journal} {Chem. Mater.}\ }\textbf {\bibinfo {volume} {25}},\ \bibinfo {pages} {555} (\bibinfo {year} {2013})},\ \bibinfo {note} {publisher: American Chemical Society}\BibitemShut {NoStop}%
\bibitem [{\citenamefont {Tong}\ \emph {et~al.}(2017)\citenamefont {Tong}, \citenamefont {Walsh}, \citenamefont {Mulvaney}, \citenamefont {Etheridge},\ and\ \citenamefont {Funston}}]{tong_control_2017}%
  \BibitemOpen
  \bibfield  {author} {\bibinfo {author} {\bibfnamefont {W.}~\bibnamefont {Tong}}, \bibinfo {author} {\bibfnamefont {M.~J.}\ \bibnamefont {Walsh}}, \bibinfo {author} {\bibfnamefont {P.}~\bibnamefont {Mulvaney}}, \bibinfo {author} {\bibfnamefont {J.}~\bibnamefont {Etheridge}},\ and\ \bibinfo {author} {\bibfnamefont {A.~M.}\ \bibnamefont {Funston}},\ }\bibfield  {title} {\bibinfo {title} {Control of {Symmetry} {Breaking} {Size} and {Aspect} {Ratio} in {Gold} {Nanorods}: {Underlying} {Role} of {Silver} {Nitrate}},\ }\href {https://doi.org/10.1021/acs.jpcc.6b10343} {\bibfield  {journal} {\bibinfo  {journal} {J. Phys. Chem. C}\ }\textbf {\bibinfo {volume} {121}},\ \bibinfo {pages} {3549} (\bibinfo {year} {2017})},\ \bibinfo {note} {publisher: American Chemical Society}\BibitemShut {NoStop}%
\bibitem [{\citenamefont {Takenaka}\ \emph {et~al.}(2013)\citenamefont {Takenaka}, \citenamefont {Kawabata}, \citenamefont {Kitahata}, \citenamefont {Yoshida}, \citenamefont {Matsuzawa},\ and\ \citenamefont {Ohzono}}]{takenaka_effects_2013}%
  \BibitemOpen
  \bibfield  {author} {\bibinfo {author} {\bibfnamefont {Y.}~\bibnamefont {Takenaka}}, \bibinfo {author} {\bibfnamefont {Y.}~\bibnamefont {Kawabata}}, \bibinfo {author} {\bibfnamefont {H.}~\bibnamefont {Kitahata}}, \bibinfo {author} {\bibfnamefont {M.}~\bibnamefont {Yoshida}}, \bibinfo {author} {\bibfnamefont {Y.}~\bibnamefont {Matsuzawa}},\ and\ \bibinfo {author} {\bibfnamefont {T.}~\bibnamefont {Ohzono}},\ }\bibfield  {title} {\bibinfo {title} {Effects of surfactant concentration on formation of high-aspect-ratio gold nanorods},\ }\href {https://doi.org/10.1016/j.jcis.2013.06.008} {\bibfield  {journal} {\bibinfo  {journal} {Journal of Colloid and Interface Science}\ }\textbf {\bibinfo {volume} {407}},\ \bibinfo {pages} {265} (\bibinfo {year} {2013})}\BibitemShut {NoStop}%
\bibitem [{\citenamefont {Walsh}\ \emph {et~al.}(2017)\citenamefont {Walsh}, \citenamefont {Tong}, \citenamefont {Katz-Boon}, \citenamefont {Mulvaney}, \citenamefont {Etheridge},\ and\ \citenamefont {Funston}}]{walsh_mechanism_2017}%
  \BibitemOpen
  \bibfield  {author} {\bibinfo {author} {\bibfnamefont {M.~J.}\ \bibnamefont {Walsh}}, \bibinfo {author} {\bibfnamefont {W.}~\bibnamefont {Tong}}, \bibinfo {author} {\bibfnamefont {H.}~\bibnamefont {Katz-Boon}}, \bibinfo {author} {\bibfnamefont {P.}~\bibnamefont {Mulvaney}}, \bibinfo {author} {\bibfnamefont {J.}~\bibnamefont {Etheridge}},\ and\ \bibinfo {author} {\bibfnamefont {A.~M.}\ \bibnamefont {Funston}},\ }\bibfield  {title} {\bibinfo {title} {A {Mechanism} for {Symmetry} {Breaking} and {Shape} {Control} in {Single}-{Crystal} {Gold} {Nanorods}},\ }\href {https://doi.org/10.1021/acs.accounts.7b00313} {\bibfield  {journal} {\bibinfo  {journal} {Acc. Chem. Res.}\ }\textbf {\bibinfo {volume} {50}},\ \bibinfo {pages} {2925} (\bibinfo {year} {2017})},\ \bibinfo {note} {publisher: American Chemical Society}\BibitemShut {NoStop}%
\bibitem [{\citenamefont {Xiao}\ and\ \citenamefont {Qi}(2011)}]{xiao_surfactant_2011}%
  \BibitemOpen
  \bibfield  {author} {\bibinfo {author} {\bibfnamefont {J.}~\bibnamefont {Xiao}}\ and\ \bibinfo {author} {\bibfnamefont {L.}~\bibnamefont {Qi}},\ }\bibfield  {title} {\bibinfo {title} {Surfactant -assisted, shape-controlled synthesis of gold nanocrystals},\ }\href {https://doi.org/10.1039/C0NR00814A} {\bibfield  {journal} {\bibinfo  {journal} {Nanoscale}\ }\textbf {\bibinfo {volume} {3}},\ \bibinfo {pages} {1383} (\bibinfo {year} {2011})},\ \bibinfo {note} {publisher: Royal Society of Chemistry}\BibitemShut {NoStop}%
\bibitem [{\citenamefont {Gao}\ \emph {et~al.}(2003)\citenamefont {Gao}, \citenamefont {Bender},\ and\ \citenamefont {Murphy}}]{gao_dependence_2003}%
  \BibitemOpen
  \bibfield  {author} {\bibinfo {author} {\bibfnamefont {J.}~\bibnamefont {Gao}}, \bibinfo {author} {\bibfnamefont {C.~M.}\ \bibnamefont {Bender}},\ and\ \bibinfo {author} {\bibfnamefont {C.~J.}\ \bibnamefont {Murphy}},\ }\bibfield  {title} {\bibinfo {title} {Dependence of the {Gold} {Nanorod} {Aspect} {Ratio} on the {Nature} of the {Directing} {Surfactant} in {Aqueous} {Solution}},\ }\href {https://doi.org/10.1021/la034919i} {\bibfield  {journal} {\bibinfo  {journal} {Langmuir}\ }\textbf {\bibinfo {volume} {19}},\ \bibinfo {pages} {9065} (\bibinfo {year} {2003})},\ \bibinfo {note} {publisher: American Chemical Society}\BibitemShut {NoStop}%
\bibitem [{\citenamefont {Sekerka}(1993)}]{sekerka_role_1993}%
  \BibitemOpen
  \bibfield  {author} {\bibinfo {author} {\bibfnamefont {R.~F.}\ \bibnamefont {Sekerka}},\ }\bibfield  {title} {\bibinfo {title} {Role of instabilities in determination of the shapes of growing crystals},\ }\href {https://doi.org/10.1016/0022-0248(93)90288-8} {\bibfield  {journal} {\bibinfo  {journal} {Journal of Crystal Growth}\ }\textbf {\bibinfo {volume} {128}},\ \bibinfo {pages} {1} (\bibinfo {year} {1993})}\BibitemShut {NoStop}%
\bibitem [{\citenamefont {Marks}\ and\ \citenamefont {Peng}(2016)}]{marks_nanoparticle_2016}%
  \BibitemOpen
  \bibfield  {author} {\bibinfo {author} {\bibfnamefont {L.~D.}\ \bibnamefont {Marks}}\ and\ \bibinfo {author} {\bibfnamefont {L.}~\bibnamefont {Peng}},\ }\bibfield  {title} {\bibinfo {title} {Nanoparticle shape, thermodynamics and kinetics},\ }\href {https://doi.org/10.1088/0953-8984/28/5/053001} {\bibfield  {journal} {\bibinfo  {journal} {J. Phys.: Condens. Matter}\ }\textbf {\bibinfo {volume} {28}},\ \bibinfo {pages} {053001} (\bibinfo {year} {2016})},\ \bibinfo {note} {publisher: IOP Publishing}\BibitemShut {NoStop}%
\bibitem [{\citenamefont {da~Silva}\ \emph {et~al.}(2025)\citenamefont {da~Silva}, \citenamefont {Netz},\ and\ \citenamefont {Meneghetti}}]{da_silva_applying_2025}%
  \BibitemOpen
  \bibfield  {author} {\bibinfo {author} {\bibfnamefont {J.~A.}\ \bibnamefont {da~Silva}}, \bibinfo {author} {\bibfnamefont {P.~A.}\ \bibnamefont {Netz}},\ and\ \bibinfo {author} {\bibfnamefont {M.~R.}\ \bibnamefont {Meneghetti}},\ }\bibfield  {title} {\bibinfo {title} {Applying {Molecular} {Dynamics} {Simulations} to {Unveil} the {Anisotropic} {Growth} {Mechanism} of {Gold} {Nanorods}: {Advances} and {Perspectives}},\ }\bibfield  {journal} {\bibinfo  {journal} {J. Chem. Inf. Model.}\ }\href {https://doi.org/10.1021/acs.jcim.4c02009} {10.1021/acs.jcim.4c02009} (\bibinfo {year} {2025}),\ \bibinfo {note} {publisher: American Chemical Society}\BibitemShut {NoStop}%
\bibitem [{\citenamefont {Bassani}\ and\ \citenamefont {Engel}(2025)}]{bassani_kinetically_2025}%
  \BibitemOpen
  \bibfield  {author} {\bibinfo {author} {\bibfnamefont {C.~L.}\ \bibnamefont {Bassani}}\ and\ \bibinfo {author} {\bibfnamefont {M.}~\bibnamefont {Engel}},\ }\bibfield  {title} {\bibinfo {title} {Kinetically {Trapped} {Nanocrystals} with {Symmetry}-{Preserving} {Shapes}},\ }\bibfield  {journal} {\bibinfo  {journal} {J. Am. Chem. Soc.}\ }\href {https://doi.org/10.1021/jacs.4c17157} {10.1021/jacs.4c17157} (\bibinfo {year} {2025}),\ \bibinfo {note} {publisher: American Chemical Society}\BibitemShut {NoStop}%
\bibitem [{\citenamefont {Balankura}\ \emph {et~al.}(2016)\citenamefont {Balankura}, \citenamefont {Qi}, \citenamefont {Zhou},\ and\ \citenamefont {Fichthorn}}]{balankura_predicting_2016}%
  \BibitemOpen
  \bibfield  {author} {\bibinfo {author} {\bibfnamefont {T.}~\bibnamefont {Balankura}}, \bibinfo {author} {\bibfnamefont {X.}~\bibnamefont {Qi}}, \bibinfo {author} {\bibfnamefont {Y.}~\bibnamefont {Zhou}},\ and\ \bibinfo {author} {\bibfnamefont {K.~A.}\ \bibnamefont {Fichthorn}},\ }\bibfield  {title} {\bibinfo {title} {Predicting kinetic nanocrystal shapes through multi-scale theory and simulation: {Polyvinylpyrrolidone}-mediated growth of {Ag} nanocrystals},\ }\href {https://doi.org/10.1063/1.4964297} {\bibfield  {journal} {\bibinfo  {journal} {The Journal of Chemical Physics}\ }\textbf {\bibinfo {volume} {145}},\ \bibinfo {pages} {144106} (\bibinfo {year} {2016})}\BibitemShut {NoStop}%
\bibitem [{\citenamefont {Seyed-Razavi}\ \emph {et~al.}(2011)\citenamefont {Seyed-Razavi}, \citenamefont {Snook},\ and\ \citenamefont {Barnard}}]{seyed-razavi_surface_2011}%
  \BibitemOpen
  \bibfield  {author} {\bibinfo {author} {\bibfnamefont {A.}~\bibnamefont {Seyed-Razavi}}, \bibinfo {author} {\bibfnamefont {I.~K.}\ \bibnamefont {Snook}},\ and\ \bibinfo {author} {\bibfnamefont {A.~S.}\ \bibnamefont {Barnard}},\ }\bibfield  {title} {\bibinfo {title} {Surface {Area} {Limited} {Model} for {Predicting} {Anisotropic} {Coarsening} of {Faceted} {Nanoparticles}},\ }\href {https://doi.org/10.1021/cg101088d} {\bibfield  {journal} {\bibinfo  {journal} {Crystal Growth \& Design}\ }\textbf {\bibinfo {volume} {11}},\ \bibinfo {pages} {158} (\bibinfo {year} {2011})},\ \bibinfo {note} {publisher: American Chemical Society}\BibitemShut {NoStop}%
\bibitem [{\citenamefont {Xia}\ \emph {et~al.}(2015)\citenamefont {Xia}, \citenamefont {Xia},\ and\ \citenamefont {Peng}}]{xia_shape-controlled_2015}%
  \BibitemOpen
  \bibfield  {author} {\bibinfo {author} {\bibfnamefont {Y.}~\bibnamefont {Xia}}, \bibinfo {author} {\bibfnamefont {X.}~\bibnamefont {Xia}},\ and\ \bibinfo {author} {\bibfnamefont {H.-C.}\ \bibnamefont {Peng}},\ }\bibfield  {title} {\bibinfo {title} {Shape-{Controlled} {Synthesis} of {Colloidal} {Metal} {Nanocrystals}: {Thermodynamic} versus {Kinetic} {Products}},\ }\href {https://doi.org/10.1021/jacs.5b04641} {\bibfield  {journal} {\bibinfo  {journal} {J. Am. Chem. Soc.}\ }\textbf {\bibinfo {volume} {137}},\ \bibinfo {pages} {7947} (\bibinfo {year} {2015})},\ \bibinfo {note} {publisher: American Chemical Society}\BibitemShut {NoStop}%
\bibitem [{\citenamefont {Gillespie}(2007)}]{gillespie_stochastic_2007}%
  \BibitemOpen
  \bibfield  {author} {\bibinfo {author} {\bibfnamefont {D.~T.}\ \bibnamefont {Gillespie}},\ }\bibfield  {title} {\bibinfo {title} {Stochastic {Simulation} of {Chemical} {Kinetics}},\ }\href {https://www.annualreviews.org/content/journals/10.1146/annurev.physchem.58.032806.104637} {\bibfield  {journal} {\bibinfo  {journal} {Annual Review of Physical Chemistry}\ }\textbf {\bibinfo {volume} {58}},\ \bibinfo {pages} {35} (\bibinfo {year} {2007})}\BibitemShut {NoStop}%
\bibitem [{\citenamefont {Joswiak}\ \emph {et~al.}(2016)\citenamefont {Joswiak}, \citenamefont {Doherty},\ and\ \citenamefont {Peters}}]{joswiak_critical_2016}%
  \BibitemOpen
  \bibfield  {author} {\bibinfo {author} {\bibfnamefont {M.~N.}\ \bibnamefont {Joswiak}}, \bibinfo {author} {\bibfnamefont {M.~F.}\ \bibnamefont {Doherty}},\ and\ \bibinfo {author} {\bibfnamefont {B.}~\bibnamefont {Peters}},\ }\bibfield  {title} {\bibinfo {title} {Critical length of a one-dimensional nucleus},\ }\href {https://doi.org/10.1063/1.4962448} {\bibfield  {journal} {\bibinfo  {journal} {The Journal of Chemical Physics}\ }\textbf {\bibinfo {volume} {145}},\ \bibinfo {pages} {211916} (\bibinfo {year} {2016})}\BibitemShut {NoStop}%
\bibitem [{\citenamefont {Ye}\ \emph {et~al.}(2016)\citenamefont {Ye}, \citenamefont {Jones}, \citenamefont {Frechette}, \citenamefont {Chen}, \citenamefont {Powers}, \citenamefont {Ercius}, \citenamefont {Dunn}, \citenamefont {Rotskoff}, \citenamefont {Nguyen}, \citenamefont {Adiga}, \citenamefont {Zettl}, \citenamefont {Rabani}, \citenamefont {Geissler},\ and\ \citenamefont {Alivisatos}}]{ye_single-particle_2016}%
  \BibitemOpen
  \bibfield  {author} {\bibinfo {author} {\bibfnamefont {X.}~\bibnamefont {Ye}}, \bibinfo {author} {\bibfnamefont {M.~R.}\ \bibnamefont {Jones}}, \bibinfo {author} {\bibfnamefont {L.~B.}\ \bibnamefont {Frechette}}, \bibinfo {author} {\bibfnamefont {Q.}~\bibnamefont {Chen}}, \bibinfo {author} {\bibfnamefont {A.~S.}\ \bibnamefont {Powers}}, \bibinfo {author} {\bibfnamefont {P.}~\bibnamefont {Ercius}}, \bibinfo {author} {\bibfnamefont {G.}~\bibnamefont {Dunn}}, \bibinfo {author} {\bibfnamefont {G.~M.}\ \bibnamefont {Rotskoff}}, \bibinfo {author} {\bibfnamefont {S.~C.}\ \bibnamefont {Nguyen}}, \bibinfo {author} {\bibfnamefont {V.~P.}\ \bibnamefont {Adiga}}, \bibinfo {author} {\bibfnamefont {A.}~\bibnamefont {Zettl}}, \bibinfo {author} {\bibfnamefont {E.}~\bibnamefont {Rabani}}, \bibinfo {author} {\bibfnamefont {P.~L.}\ \bibnamefont {Geissler}},\ and\ \bibinfo {author} {\bibfnamefont {A.~P.}\ \bibnamefont {Alivisatos}},\ }\bibfield  {title} {\bibinfo {title} {Single-particle mapping of nonequilibrium nanocrystal
  transformations},\ }\href {https://doi.org/10.1126/science.aah4434} {\bibfield  {journal} {\bibinfo  {journal} {Science}\ }\textbf {\bibinfo {volume} {354}},\ \bibinfo {pages} {874} (\bibinfo {year} {2016})},\ \bibinfo {note} {publisher: American Association for the Advancement of Science}\BibitemShut {NoStop}%
\bibitem [{\citenamefont {Lai}\ \emph {et~al.}(2023)\citenamefont {Lai}, \citenamefont {Liu},\ and\ \citenamefont {Evans}}]{lai_nucleation-mediated_2023}%
  \BibitemOpen
  \bibfield  {author} {\bibinfo {author} {\bibfnamefont {K.~C.}\ \bibnamefont {Lai}}, \bibinfo {author} {\bibfnamefont {D.-J.}\ \bibnamefont {Liu}},\ and\ \bibinfo {author} {\bibfnamefont {J.~W.}\ \bibnamefont {Evans}},\ }\bibfield  {title} {\bibinfo {title} {Nucleation-mediated reshaping of facetted metallic nanocrystals: {Breakdown} of the classical free energy picture},\ }\href {https://doi.org/10.1063/5.0138266} {\bibfield  {journal} {\bibinfo  {journal} {The Journal of Chemical Physics}\ }\textbf {\bibinfo {volume} {158}},\ \bibinfo {pages} {104102} (\bibinfo {year} {2023})}\BibitemShut {NoStop}%
\bibitem [{\citenamefont {Lai}\ and\ \citenamefont {Evans}(2019)}]{lai_reshaping_2019}%
  \BibitemOpen
  \bibfield  {author} {\bibinfo {author} {\bibfnamefont {K.~C.}\ \bibnamefont {Lai}}\ and\ \bibinfo {author} {\bibfnamefont {J.~W.}\ \bibnamefont {Evans}},\ }\bibfield  {title} {\bibinfo {title} {Reshaping and sintering of {3D} fcc metal nanoclusters: {Stochastic} atomistic modeling with realistic surface diffusion kinetics},\ }\href {https://doi.org/10.1103/PhysRevMaterials.3.026001} {\bibfield  {journal} {\bibinfo  {journal} {Phys. Rev. Materials}\ }\textbf {\bibinfo {volume} {3}},\ \bibinfo {pages} {026001} (\bibinfo {year} {2019})}\BibitemShut {NoStop}%
\bibitem [{\citenamefont {Han}\ \emph {et~al.}(2014)\citenamefont {Han}, \citenamefont {Liu},\ and\ \citenamefont {Evans}}]{han_real-time_2014}%
  \BibitemOpen
  \bibfield  {author} {\bibinfo {author} {\bibfnamefont {Y.}~\bibnamefont {Han}}, \bibinfo {author} {\bibfnamefont {D.-J.}\ \bibnamefont {Liu}},\ and\ \bibinfo {author} {\bibfnamefont {J.~W.}\ \bibnamefont {Evans}},\ }\bibfield  {title} {\bibinfo {title} {Real-{Time} {Ab} {Initio} {KMC} {Simulation} of the {Self}-{Assembly} and {Sintering} of {Bimetallic} {Epitaxial} {Nanoclusters}: {Au} + {Ag} on {Ag}(100)},\ }\href {https://doi.org/10.1021/nl5017128} {\bibfield  {journal} {\bibinfo  {journal} {Nano Lett.}\ }\textbf {\bibinfo {volume} {14}},\ \bibinfo {pages} {4646} (\bibinfo {year} {2014})},\ \bibinfo {note} {publisher: American Chemical Society}\BibitemShut {NoStop}%
\bibitem [{\citenamefont {Turner}\ \emph {et~al.}(2016)\citenamefont {Turner}, \citenamefont {Lei},\ and\ \citenamefont {Bao}}]{turner_modeling_2016}%
  \BibitemOpen
  \bibfield  {author} {\bibinfo {author} {\bibfnamefont {C.~H.}\ \bibnamefont {Turner}}, \bibinfo {author} {\bibfnamefont {Y.}~\bibnamefont {Lei}},\ and\ \bibinfo {author} {\bibfnamefont {Y.}~\bibnamefont {Bao}},\ }\bibfield  {title} {\bibinfo {title} {Modeling the atomistic growth behavior of gold nanoparticles in solution},\ }\href {https://doi.org/10.1039/C6NR01881E} {\bibfield  {journal} {\bibinfo  {journal} {Nanoscale}\ }\textbf {\bibinfo {volume} {8}},\ \bibinfo {pages} {9354} (\bibinfo {year} {2016})},\ \bibinfo {note} {publisher: The Royal Society of Chemistry}\BibitemShut {NoStop}%
\bibitem [{\citenamefont {Chandler}(1987)}]{chandler_introduction_1988}%
  \BibitemOpen
  \bibfield  {author} {\bibinfo {author} {\bibfnamefont {D.}~\bibnamefont {Chandler}},\ }\href@noop {} {\emph {\bibinfo {title} {\textit{{Introduction} to {Modern} {Statistical} {Mechanics}}}}}\ (\bibinfo  {publisher} {Oxford University Press},\ \bibinfo {address} {New York},\ \bibinfo {year} {1987})\ Chap.~\bibinfo {chapter} {5}\BibitemShut {NoStop}%
\bibitem [{\citenamefont {Theodorou}\ and\ \citenamefont {Suter}(1985)}]{theodorou_shape_1985}%
  \BibitemOpen
  \bibfield  {author} {\bibinfo {author} {\bibfnamefont {D.~N.}\ \bibnamefont {Theodorou}}\ and\ \bibinfo {author} {\bibfnamefont {U.~W.}\ \bibnamefont {Suter}},\ }\bibfield  {title} {\bibinfo {title} {Shape of unperturbed linear polymers: polypropylene},\ }\href {https://doi.org/10.1021/ma00148a028} {\bibfield  {journal} {\bibinfo  {journal} {Macromolecules}\ }\textbf {\bibinfo {volume} {18}},\ \bibinfo {pages} {1206} (\bibinfo {year} {1985})},\ \bibinfo {note} {publisher: American Chemical Society}\BibitemShut {NoStop}%
\bibitem [{\citenamefont {Chandler}(2005)}]{chandler_interfaces_2005}%
  \BibitemOpen
  \bibfield  {author} {\bibinfo {author} {\bibfnamefont {D.}~\bibnamefont {Chandler}},\ }\bibfield  {title} {\bibinfo {title} {Interfaces and the driving force of hydrophobic assembly},\ }\href {https://doi.org/10.1038/nature04162} {\bibfield  {journal} {\bibinfo  {journal} {Nature}\ }\textbf {\bibinfo {volume} {437}},\ \bibinfo {pages} {640} (\bibinfo {year} {2005})},\ \bibinfo {note} {publisher: Nature Publishing Group}\BibitemShut {NoStop}%
\bibitem [{\citenamefont {Oaks-Leaf}\ and\ \citenamefont {Limmer}()}]{oaksleaf_data}%
  \BibitemOpen
  \bibfield  {author} {\bibinfo {author} {\bibfnamefont {S.}~\bibnamefont {Oaks-Leaf}}\ and\ \bibinfo {author} {\bibfnamefont {D.~T.}\ \bibnamefont {Limmer}},\ }\href@noop {} {}\bibinfo {howpublished} {https://doi.org/10.5281/zenodo.15052663}\BibitemShut {NoStop}%
\end{thebibliography}%

\clearpage

\section*{Supplementary Material for Mechanism of Shape Symmetry Breaking in Surfactant Mediated Crystal Growth} 
 
\section{\label{sec:RateDetermination} Determination of Rates in Markov Models}

We set the rates of the Markov model presented in \textbf{Fig. 2} such that they agree with the analytical results obtained by Joswiak \textit{et. al.}\cite{joswiak_critical_2016} for a one-dimensional nucleation process. We follow the method they outline to solve for the committor of nucleation on a finite facet of side length $h_j$. We obtain
\begin{equation}
    K_s^{+} = \frac{h_j k_s^{+}(k_s^{+}-k_s^{r})}{(k_s^{+}-k_s^{r}) + k_s^{d}(1-(k_s^{r}/k_s^{+})^{h_j-1})}
    \label{GrowthRate}
\end{equation}
where $s$ could refer to species $g$ or $b$. $k_s^{+}$ is the rate of attachment for species $s$ at any lattice site. For simplicity, we set $k^{+}_g = k^{+}_b=\tau^{-1}$ throughout and measure all times in units of $\tau$. Then the detachment rates of species $s$ are given by a rate for the nucleus to recede, $k_s^{r} = \tau^{-1}\exp(-\beta\Delta\mu_s)$, and a rate for it to dissolve, $k^{d}_s=\tau^{-1}\exp(-\beta\Delta\mu_s+\beta\epsilon_s)$. Physically $k_s^{r}$ would correspond to removing an monomer from the edge of the growing nucleus, while $k_s^{d}$ is the rate to detach a single monomer with no lateral neighbors of the same kind. The factor of $h_j$ in the numerator accounts for the fact that a nucleation event could occur independently at any location on the facet. We ignore effects of multiple nucleation. Our definitions of $k_s^{+}$, $k_s^{r}$, and $k_s^{d}$ correspond exactly to the rates used in the two-component lattice gas in \textbf{Fig. 3}. Specifically, there is uniform attachment for both species with rate $\tau^{-1}$, and a detachment rate that depends on the number of nearest neighbors of the same species, which is $k_s^{-}(n_s)=\tau^{-1}\exp(-\beta\mu_s +\beta\epsilon_sn_s)$.

For the model of \textbf{Fig. 2} we include the competitive rate $K_{bg}$ in the following way. We propose that a nucleus of species $g$ can form at the edge of a layer of species $b$ if a monomer of species $b$ is removed and the empty site is filled by species $g$ before it can be filled by species $b$ again. We assign the rate for this concerted process as $k_{bg}^{+} = \frac{k_b^{r}k_g^{+}}{k_b^{+} + k_g^{+}}=\tau^{-1}\exp(-\beta\Delta\mu_b)/2$. Similarly, one sees that for this concerted process, we have $k_{bg}^{d} = \frac{k_b^{+}k_g^{d}}{k_b^{+} + k_g^{+}}=\tau^{-1}\exp(-\beta\Delta\mu_g +\beta\epsilon_g)/2$, and $k_{bg}^{r} = \frac{k_b^{+}k_g^{r}}{k_b^{+} + k_g^{+}}=\tau^{-1}\exp(-\beta\Delta\mu_g)/2$. We plug in these rates to equation \eqref{GrowthRate} to obtain $K_{bg}$,
\begin{equation}
K_{bg}= \frac{ k_{bg}^{+}(k_{bg}^{+}-k_{bg}^{r})}{(k_{bg}^{+}-k_{bg}^{r}) + k_{bg}^{d}(1-(k_{bg}^{r}/k_{bg}^{+})^{h_j-1})}
\end{equation}
note that we remove the factor of $h_j$ in the numerator because this process occurs only at the edge of the facet. An analogous process could occur at the center of the facet, but, at least under the assumptions of the coarse-grained model, this process would be driven by a rate $k^{+}_{bg,\mathrm{center}}=\tau^{-1}\exp(-\beta\Delta\mu_b -\beta\epsilon_b)/2$. We use values of $\beta\epsilon_b$ on the order of $\sim 15$, thus the facet would need to be millions of monomers in length for this rate to contribute. 

\section{\label{sec:UnattainedTrials}Simulation Details for two component lattice gas}
The trajectories of \textbf{Fig. 3b} were all intended to run until the system reached a total size of 2000 monomers. Due to computational constraints, simulations were cutoff once they reached $10^{10}$ KMC steps, and this cutoff caused some configurations not to reach the desired size. Under conditions $\beta\Delta\mu_b = 1.5$, $1.6$, and $1.7$, there were respectively $7$, $38$, and $20$ trials out of 1024 total trials that did not reach the desired size. These configurations are not included in the histogram of \textbf{Fig. 3b}. It was additionally observed in some of the trials that did not attain the final size that there were multiple distinct clusters of metal monomers, rather than a single crystal. This can occur when monomers are removed such that a cluster detaches from the initial seed. These configurations are unphysical since our model assumes a single growing seed and does not attempt to represent the surrounding fluid or effects of other seeds in any way. However, they are also irrelevant to purposes of this letter in establishing the existence of a mechanism for anisotropic growth. One can view the final configurations of trials that did not reach the final size in \textbf{Supplementary Videos 1-3}.
\\
\\
\textbf{Supplementary Video 1} Final configurations of trials that did not attain the desired size in \textbf{Fig. 3b.} with $\beta\Delta\mu_b = 1.5$.
\\
\\
\textbf{Supplementary Video 2} Final configurations of trials that did not attain the desired size in \textbf{Fig. 3b.} with $\beta\Delta\mu_b = 1.6$.
\\
\\
\textbf{Supplementary Video 3} Final configurations of trials that did not attain the desired size in \textbf{Fig. 3b.} with $\beta\Delta\mu_b = 1.7$.

\end{document}